# ELM-free Enhanced $D_\alpha$ H-mode with Near Zero NBI Torque Injection in DIII-D Tokamak

T. Macwan[1, *], K. Barada[1], J. F. Parisi[2], R. Groebner[3], T. L. Rhodes[1], S. Banerjee[2], C. Chrystal[3], Q. Pratt[1], Z. Yan[4], H. Wang[3], L. Zeng[1], M. E. Austin[5], N. A. Crocker[1], and W. A. Peebles[1]

[1] Physics and Astronomy Department, PO Box 957099, Los Angeles, CA 90095-7099, United States of America

[2] Princeton Plasma Physics Laboratory, PO Box 451, Princeton, NJ 08543-0451, United States of America

[3] General Atomics, PO Box 85608, San Diego, CA 92186-5608, United States of America

[4] University of Wisconsin-Madison, 1500 Engineering Dr., Madison, Wisconsin 53706, United States of America

[5] The University of Texas at Austin, Austin, Texas 78712, USA

*Email: tmacwan@physics.ucla.edu

## Abstract

Enhanced $D_\alpha$ H-mode (EDA H-mode), an ELM-free H-mode regime, is explored in neutral beam heated, lower single null plasmas with near zero torque injection. This regime exhibits a good energy confinement ($H_{98y2} \sim 1$) with $\beta_N \sim 2$, high density, regime access at low input power, and no ELMs. This paper further presents the time–resolved measurements of electron and ion density, temperature, plasma rotation, and radial electric field during the EDA H-mode phase and examines the dynamics of the edge quasi-coherent mode (QCM). Measurements using multiple fluctuation diagnostics reveal the QCM to be a separatrix spanning mode, peaking just inside the separatrix, existing in a wide range of $k_\perp \rho_s \sim 0.1 - 1.2$ with multiple harmonics, and propagating with a very small phase velocity in the plasma frame, where $k_\perp$ is the binormal wavenumber and $\rho_s$ is the ion sound radius. Linear gyrokinetic simulations of an EDA H-mode discharge with CGYRO indicates that the trapped electron mode (TEM) and electron temperature gradient (ETG) are dominant instabilities in the region where QCM is unstable. Qualitative analysis indicates that the properties of TEM are consistent with the experimental observed characteristics of the QCM. These similarities suggest that the QCM might be a TEM instability existing in the edge region of the EDA H-mode plasmas.

## 1. Introduction

H-mode [1] is an attractive choice for the operation of future fusion devices due to its improved energy confinement. However, an H-mode is usually accompanied by intermittent edge localized modes (ELMs) [2], which limit the maximum achievable pedestal pressure [3]. Although ELMs lead to periodic expulsion of impurities from the plasma, high transient heat flux due to large ELMs are envisioned to cause considerable damage to the divertor and wall components in ITER [4, 5] and future fusion reactors. One of the solutions is to mitigate the ELMs using techniques such as resonant magnetic perturbations (RMPs) or pellet injection [6, 7]. Alternatively, development of naturally ELM-free regimes [8] such as quiescent H-mode (QH mode) [9], wide pedestal QH (WPQH) mode [10], I-mode [11], and EDA H-mode [12] is an attractive option. Further development of these regimes is essential for the safe operation of next generation devices.

Enhanced $D_\alpha$ H-mode or EDA H-mode, originally named due to high divertor $D_\alpha$ levels, is an ELM-free regime which was discovered and subsequently extensively explored in Alcator C-mod [12–19] using ion cyclotron resonance frequency (ICRF) heating. The EDA H-mode has been subsequently studied in DIII-D [20], EAST [21] and ASDEX-Upgrade (AUG) [22, 23]. The defining feature of the EDA H-mode is an edge instability called quasi-coherent mode (QCM) which is observed in several fluctuation diagnostics. The QCM is believed to enhance the particle transport and maintain the pedestal gradients below the coupled peeling-ballooning mode boundary in EDA H-mode, hence avoiding ELMs [15, 17, 24]. QCM is also believed to flush impurities and thus maintain stationary, ELM-free discharges. This has led to high performance discharges, notably in Alcator C-Mod where highest volume-averaged pressure has been achieved to date [16]. Recently, experiments in AUG [22] demonstrated several attractive features of EDA H-mode which are desirable in fusion reactors such as possibility of access at low input power, dominant electron heating with good confinement, low impurity content with compatibility to tungsten divertor and first wall and an absence of ELMs. Subsequently, this regime was integrated with argon (Ar) seeding for radiative power removal in the pedestal with combined electron cyclotron heating (ECH) and neutral beam injection (NBI) heating of up to 5 MW [23]. These advances make EDA H-mode a very promising regime for future devices; although, its extension to low collisionality operation is still an area of active research.

The QCM essentially regulates the fuel particle and impurity transport in the EDA H-mode and has been studied extensively in Alcator C-Mod [17, 18, 19]. QCM has been measured in several fluctuation diagnostics, including phase-contrast imaging [25], reflectometry [20, 22, 26], beam emission spectroscopy [27], electrostatic and magnetic probes [22, 17], and gas-puff imaging [18]. The measurements have shown QCM to be a separatrix spanning, field-aligned perturbation, localized near radial electric field ($E_r$) well, and which drives outward radial plasma flux. The exact nature of the QCM, however, is still debated. QCM was initially identified as a resistive-ballooning, x-point mode using BOUT modeling [25] and was shown to propagate in electron diamagnetic direction in the laboratory frame experimentally as well as in modeling. Drift-Alfven waves, propagating in electron diamagnetic direction in plasma frame, were also shown as a candidate for QCM [28]. Modeling also showed QCM to be pressure driven surface waves [29]. Extensive measurements using Langmuir probes demonstrated QCM to be an electron drift wave with interchange and electromagnetic contributions, propagating in the electron diamagnetic direction in plasma frame [17]. However, recent measurement of QCM with gas puff imaging along with radial electric field measurement with gas puff charge exchange recombination spectroscopy showed that the QCM is centered near $E_r$ well minimum and propagates along the ion diamagnetic direction in plasma frame [18]. Gyrokinetic analysis of the Ar-seeded EDA H-mode discharge obtained in AUG was performed for the first-time using GENE [30] for core and pedestal locations. However, these electrostatic simulations were unable to capture the physics governing the QCMs owing to the limitation of performing local electrostatic gyrokinetic simulations in the pedestal. The non-linear magnetohydrodynamics (MHD) simulations were also performed for Ar-seeded H-mode discharge using the visco-resistive extended MHD code, JOREK [31]. The modes observed in this simulation had poloidal wavenumbers between $k_\theta \sim 0.1$–$0.5$ cm$^{-1}$ which were lower than the expected QCM wavenumber expected. These modes were shown to cause non-negligible heat and particle transport.

This paper reports on the study of EDA H-mode in the NBI heated discharges of DIII-D in lower single null plasmas with near zero net torque conditions. The paper is mainly focused on the physics of the EDA H-mode, particularly on the dynamics of the QCM. Several fluctuations diagnostics are used to measure the QCM. Notably, Doppler backscattering (DBS) diagnostic is used to measure the fine structure of QCM at high wavenumbers, its mode structure and the time evolution of its propagation velocity in plasma frame. Using the experimentally measured mode structure of the QCM, a linear gyrokinetic simulation is performed during the stationary

EDA H-mode phase to identify the dominant instability at several locations within the pedestal. Section 2 details the plasma conditions and the diagnostics used, section 3 discusses experimental observations of the EDA H-mode and QCM dynamics, section 4 details the pedestal stability and some transport characteristics of EDA H-mode, linear gyrokinetic simulations are discussed in section 5, and section 6 presents the summary of the paper.

## 2. Experimental conditions and diagnostics used

The experiments in which EDA H-mode is observed are performed with a plasma current of 1 MA at two different toroidal magnetic field strength ($B_T$ = -1.4 and -2.0 T) in DIII-D tokamak. The EDA H-mode is obtained in high triangularity, lower single null plasmas with favorable geometry, i.e. the ion $\nabla B$ drift towards the X-point. These plasmas have lower triangularity $\delta_{low}$ = 0.7 – 0.8 and upper triangularity $\delta_{up}$ ~ 0.35 as shown in Figure 1a. Further, the EDA H-mode is observed in the NBI heated deuterium discharges with near zero input torque, which are achieved using balanced co- and counter-current, tangentially injected deuterium neutral beams. The data set for NBI heated, near zero torque EDA H-mode plasmas consists of 16 discharges. The EDA-H mode discharges are observed with input torque values between -0.5 – 0.5 Nm and NBI power of 0.9 – 3.5 MW. Here, the positive torque values correspond to the co-current direction. The EDA H-mode phase is observed with input torque as low as 0.1 N-m. It is also observed at higher input torque (~ 0.75 N-m) and NBI power (4.5 MW) values briefly before it transitions to ELMy H-mode. The line-averaged electron density in the EDA H-mode discharges remains $n_e$ ~ 2.2 – 6.0 × $10^{19}$ $m^{-3}$. The effective electron pedestal collisionality ($v_e^*$) is ~ 1.0 – 5.0, normalized gyro-radius $\rho$*=$\rho$/a ~ 0.002, and the $q_{95}$ is ~ 3.5 – 5.0 during the EDA H-mode phase. The EDA H-mode phase is obtained up to normalized beta $\beta_N$ ~ 2, $H_{98y2}$ ~ 1, and Greenwald fraction ~ 0.75. Further, the discharges are obtained very close to the scaled L-H threshold power required for ITER [32]. The extension of operating parameters, especially at a lower $v_e^*$ is highly desirable and will be a subject of future work.

The electron density and temperature profiles are measured with the Thomson scattering (TS) [32, 33] diagnostics. The ion density, temperature, and rotation of the fully ionized carbon is measured using a charge exchange recombination (CER) [34] system. The radial electric field is derived from the CER data using the radial force-balance equation [35]. The averaged ion/rotation profiles and electron profiles are radially adjusted to satisfy the following two criteria respectively: while the ion and rotation profiles are shifted so that, at the separatrix, the radial electric field $E_r$ = 0; the electron profiles are shifted so that the electron temperature at the separatrix matches the value determined from a 'two-point' divertor heat flux model [36]. Several fluctuation diagnostics are used for the characterization of the quasi-coherent mode (QCM) in the EDA H-mode plasmas. Langmuir probes installed in the divertor shelf are used for the measurement of the fluctuations in both, the electron as well as ion saturation current [37], which are obtained by applying voltage ramp to the probe. The beam emission spectroscopy (BES) diagnostic measures long wavelength ($k_\perp \rho_i < 1$), localized density fluctuations from $0.8 < \rho < 1.0$ ($\rho$ is the square root of the normalized toroidal flux coordinate) at the outboard midplane. Here, $k_\perp$ is the binormal wavenumber and $\rho_i$ is the ion gyroradius. BES consists of a $8 \times 7$ 2D array of channels, each channel imaging a 0.9 (radial) × 1.2 (poloidal) (cm×cm) region in the plasma [38]. BES measures Doppler-shifted H$\alpha$ and D$\alpha$ emission from collisional excited high-energy neutral beam atoms. The light intensity fluctuation can be converted to electron density fluctuation determined by atomic physics of beam atom excitation which considers of the impact from beam energy, local electron density, local electron temperature and effective charge ($Z_{eff}$). The detailed atomic physics model is described in section III of Ref. [39]. A high frequency magnetic probe [40], installed from a diagnostic port above outboard midplane, measures the low-k poloidal magnetic fluctuations from the plasma. The magnetic fluctuations are measured with a sampling rate of 2 MHz.

The localized density fluctuations (ñ) in the intermediate wave number ($k_\perp \rho_i \sim 1$) are measured using the Doppler backscattering (DBS) diagnostics [41, 42], along with the associated Doppler shifts. The diagnostic measures the spatial, temporal, and wave number resolved ñ amplitude, as well as the lab frame binormal velocity of ñ. The diagnostic has a temporal resolution of <1 $\mu s$ and spatial resolution of < 5 mm (even less in steep electron density gradient region like the pedestal). The DBS system in DIII-D [41] launches microwave beams (50 – 75 GHz) at an oblique angle to the cut-off surface with either O-mode or X-mode polarization. For these experiments, the beams were launched with X-mode polarization. As the radiation approaches the cut-off layer, the ñ from the vicinity of the cutoff layer scatters the incident wave. When the Bragg scattering condition ($\boldsymbol{k_{\tilde{n}}} = -2\boldsymbol{k_i}$) is fulfilled, the incident wave is backscattered (180°) and is collected. Here, $\boldsymbol{k_i}$ is the incident wavenumber at the cut-off location and $\boldsymbol{k_{\tilde{n}}}$ is the density fluctuation wavenumber. The amplitude of the collected signal is proportional to ñ. The probed wavenumbers ($\boldsymbol{k_{\tilde{n}}}$) and DBS scattering locations are estimated using GENRAY [43] 3D raytracing. The binormal propagation velocity (perpendicular to the equilibrium magnetic field and the flux surface normal) of ñ results in a Doppler shifted backscattered signal with a frequency of $\omega_D = k_{\tilde{n}} v_\perp$, where $v_\perp = v_{E\times B} + v_{ph}$. Here $v_{E\times B}$ is the local $E \times B$ velocity and $v_{ph}$ is the turbulence phase velocity. Generally, $v_{E\times B} \gg v_{ph}$ and hence one can estimate the local $E \times B$ velocity and its shear at different radial locations using the DBS measurements [44]. The validity of this condition for the QCM, observed during the EDA H-mode phase, is addressed in conjunction with the local $E \times B$ velocity measured with the CER diagnostics and the propagation direction of the QCM in the plasma frame is obtained in later section. The DBS diagnostics in DIII-D are installed at two toroidally separated (by 180º) locations and are termed as DBS60 [41] and DBS5 system [42]. DBS60 system has 8 channels covering a range of 55 – 75 GHz while DBS5 consists of 5 channels within a narrow band of 63.8 – 65.6 GHz.

The Doppler shifted backscattered radiation is detected by the quadrature mixers. Spectral analysis of the complex signal consisting of the in-phase (I) and the quadrature (Q) components yields the Doppler shifted spectrum of the probed density turbulence ($k_{\tilde{n}}$). The 'amplitude' and 'phase' of the backscattered electric field can also be analyzed individually. The amplitude ($A_{DBS}(t) = \sqrt{I(t)^2 + Q(t)^2}$) is related to the level of density fluctuation over a weighted wavenumber range. The DBS phase is represented as $\phi_{DBS}(t) = \tan^{-1}(Q(t)/I(t))$ and is typically representative of the DBS Doppler shift from turbulent structures and oscillatory modes [45]. The phase of the backscattered signal typically has physical contributions from [45], assuming the density fluctuations to be of low amplitude such that the multiple scattering effects can be neglected [46]:

$$\phi_{DBS}(t) = k_\perp v_{E\times B} t + \frac{k_\perp v_m}{\omega_m} \sin \omega_m t + \tilde{\phi}(t) + 2\int_0^{x_c(t)} k(x,t)dx \qquad (1)$$

Here, first term is the Doppler shift due to the equilibrium $E \times B$ flow $v_{E\times B}$, second term is due to the presence of an oscillating coherent mode $v_m$ with frequency $\omega_m$, third term includes contributions from turbulent flows $\tilde{\phi}$ due to the advection or convection of the scattering structure by large scale structures ($k<k_\perp$) and the last term is due to the path length variations encountered by the beam. The last term is typically negligible for DBS. Further details are found in reference [41]. Taking the derivative of the first term yields the instantaneous Doppler shift of the probed density turbulence ($k_{\tilde{n}}$). Doppler shift is usually computed by performing spectral analysis of the complex signal consisting of the in-phase (I) and the quadrature (Q) components which yields the Doppler shifted spectrum of the probed density turbulence ($k_{\tilde{n}}$).

Spectral analysis of DBS phase can detect coherent oscillatory modes. Later sections employ such techniques to study the dynamics of QCM during the EDA H-mode regime. Similar analysis has been used previously for the study of oscillatory, coherent modes such as GAMs in DIII-D tokamak [45].

## 3. Observations of EDA H-mode and dynamics of quasi-coherent mode in DIII-D

This section presents the characterization of EDA H-mode discharges observed in DIII-D. The discharges presented in this section are non-stationary with the line-averaged density increasing steadily. This non-stationarity allows a study of the dynamics of the quasi-coherent mode (QCM) using several diagnostics.

*(a) EDA H-mode in DIII-D:*

The Enhanced $D_\alpha$ H-mode regime is observed in DIII-D tokamak for NBI heated discharges with near zero input torque in lower single null shaped plasmas for the discharges discussed in this paper. The EDA H-mode is essentially an H-mode with the absence of periodic ELMs. One such discharge is shown in Figure 1. The plasma is lower single null and average triangularity of $\delta_{av} = 0.56$ (Figure 1a) and elongation of 1.77. The discharge has a toroidal magnetic field of –1.4 T and a plasma current of 1.0 MA. It can be seen from figure 1f that as the NBI power is increased to 1.75 MW at 2.5 s, the plasma first transitions to H-mode, evident by a decrease in the divertor $D_\alpha$ (figure 1d) level and increase in the density (figure 1b). The input torque is kept constant at near zero value of – 0.15 N-m (figure 1f). Within ~ 15 ms, the plasma undergoes another transition to an ELM-free EDA H-mode marked by a significant rise in the $D_\alpha$ level and the appearance of the characteristic quasi-coherent mode with decreasing frequency in time (figure 1h) [12]. The gas fueling is completely turned off after the L-H transition as shown in figure 1c.

The blue shaded region in Figure 1 represents the EDA H-mode phase which lasts for ~ 800 ms during the discharge even though the discharge undergoes several changes during this phase. The EDA H-mode is characterized by absence of any ELM activity as seen in the $D_\alpha$ signal (figure 1d) and the appearance of QCM (fig. 1h). The density increases during the ELM-free phase from ~ $4.2 \times 10^{19} m^{-3}$ to $7.0 \times 10^{19} m^{-3}$ ending at a Greenwald fraction of ~ 0.72. The edge safety factor, $q_{95}$, (figure 1g) rises slightly from 3.5 to 3.7 during this phase. The normalized beta and $H_{98y2}$ factor (figure 1e) increase steadily through the EDA H-mode phase to high values of ~ 2.0 and ~ 1.0 respectively. The phase spectra of the DBS (figure 1h) clearly shows a quasi-coherent mode which starts at a high frequency of ~ 80 kHz and promptly decreases to a lower frequency of as low as ~10 kHz. A closer look also shows a harmonic of the QCM at twice the QCM frequency in the phase spectra. The NBI power is increased after 2.8 s in form of pulses which are fired for a duration of 10 ms with an off time of 10 ms. These are used for measuring the density fluctuations using BES. The increased average NBI power is 2.6 MW while the input torque is maintained at the same level. At 3.14 s the NBI power as well as the input torque are increased to 4.3 MW and 0.75 N-m. The sudden increase in torque and power is marked by an abrupt increase in the $\beta_N$ value and a decrease in the frequency of the QCM. However, the EDA H-mode is maintained for the next 180 ms and exhibits a lower QCM frequency of ~ 10 kHz. The discharge further undergoes a dithering behavior and results in a single ELM at 3.41 s where the QCM disappears. The dithering behavior before an ELM is seen in the divertor $D_\alpha$ signal (for time 3300–3350 ms) in the figure 1d as an inset in red. It is to be noted that this dithering behavior observed between 3320-3415 ms is significantly different from the behavior of $D_\alpha$ signal between ~2600-2900 ms (shown as an inset in figure

1d for time 2700–2750 ms in blue). The dithering behavior is also marked by an absence of the QCM. The EDA H-mode exists briefly (~ 20 ms) before regular ELMs appear with a frequency of ~ 100 Hz as the discharge transitions to an ELMy H-mode. The ELM-free, high confinement operation ($\beta_N$ ~ 2 and $H_{98y2}$ ~ 1) in near zero torque conditions might be relevant to future devices where the input torque is predicted to be low. A better density control in these discharges will help in maintaining stationary conditions, more suitable for future studies. It is to be noted that high confinement, stationary EDA H-mode discharges have been observed in several devices [15, 16, 22, 47]. Further work needs to be carried out, particularly in achieving stationary EDA H-mode with low-collisionality pedestals, to make this regime suitable for future fusion devices.

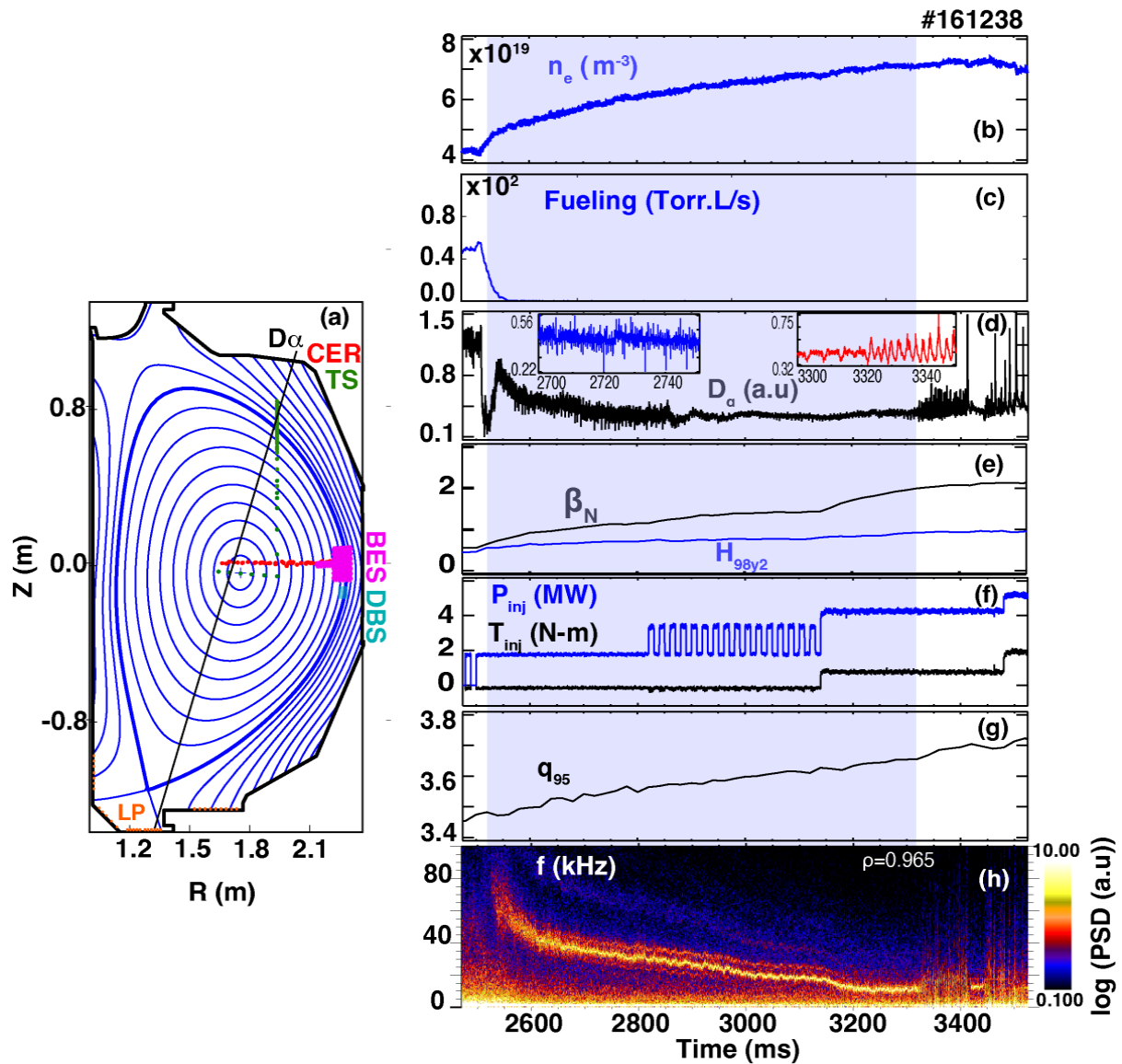

*Figure 1: EDA H-mode (shaded) discharge with: (a) lower single null plasma equilibrium with measurement locations for the main diagnostics (CER - charge exchange recombination, BES - beam emission spectroscopy, TS - Thomson scattering, LP - Langmuir probes, $D_\alpha$ - divertor $D_\alpha$ line of sight). The plotted flux surfaces depict equally spaced ρ values. Time evolution of (b) chord averaged electron density, (c) $D_2$ gas fueling (d) divertor $D_\alpha$, (e) $\beta_N$ (black) and $H_{98y2}$ (blue) factor, (f) injected NBI power (blue) and NBI torque (black), (g) $q_{95}$, and (h) DBS phase spectra (DBS8, X-mode, 62.5 GHz,) showing the QCM during EDA H-mode regime. Two insets in (c) show the divertor $D_\alpha$ signal for times (2700–2750 ms, in blue) and (3300–3350 ms, in red).*

*(b) Observation of QCM in multiple diagnostics*

The existence of ELM-free EDA H-mode is synonymous with the presence of an edge fluctuation known as the quasi-coherent mode (QCM) [12]. In DIII-D discharges, the QCM also appears in several fluctuation diagnostics during the EDA H-mode phase such as the Langmuir probes, magnetic probes, and beam emission spectroscopy. In this work, in addition to the observation of QCM in these diagnostics, the QCM is studied in detail with the intermediate-k fluctuation diagnostics of Doppler Backscattering (DBS). As discussed in section 2, the DBS diagnostics used in this experiment can measure the density fluctuations in the range of $k_\perp \sim 1 - 10$ cm$^{-1}$, with $k_\perp \rho_i \sim 1$ in the pedestal region.

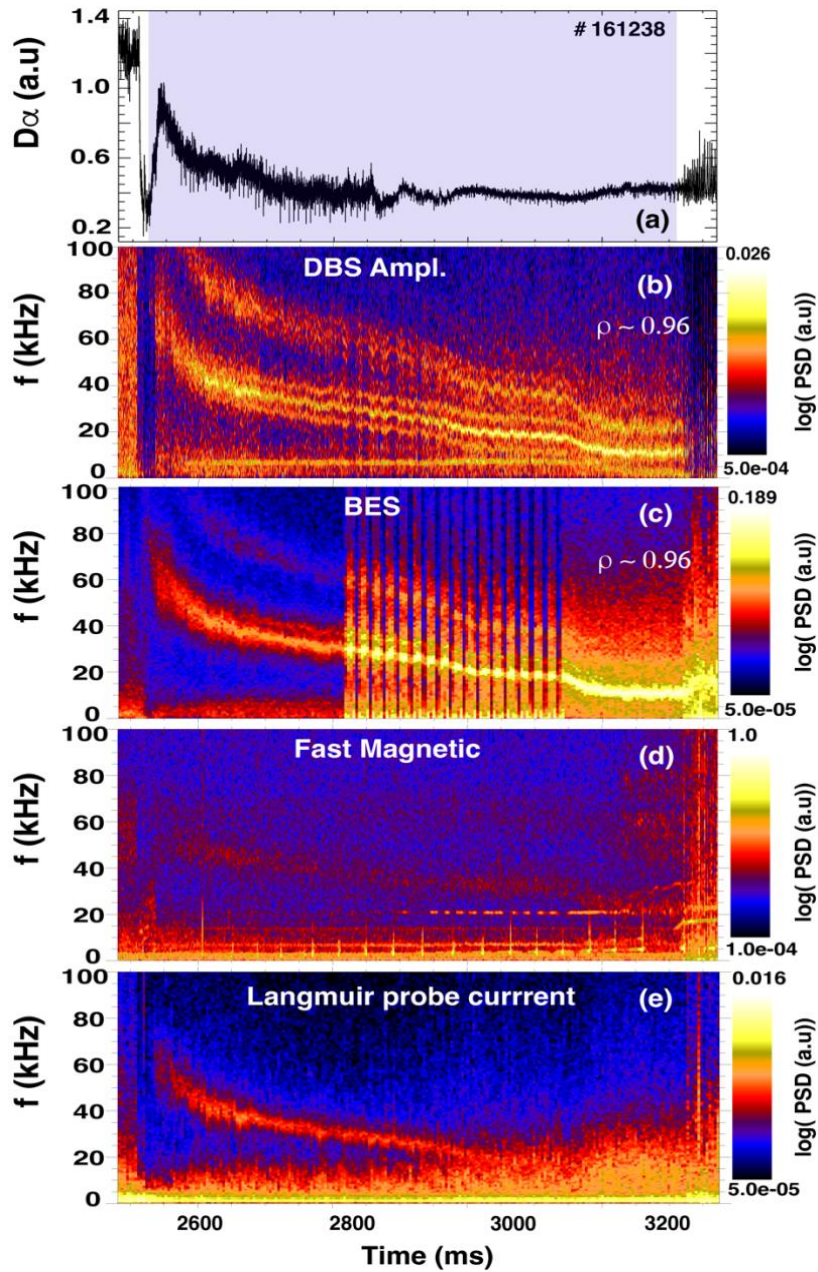

*Figure 2: Observation of QCM during EDA H-mode phase (shaded region with elevated $D_\alpha$ (a)) in multiple diagnostics: (b) DBS amplitude spectra, (c) BES spectra, (d) fast magnetics spectra, and (e) current spectrum of Langmuir probe. The observed QCM frequency as well as its change with time is very similar for all diagnostics.*

The QCM is observed during the EDA H-mode in several diagnostics as shown in figure 5. The vertical dashed line in Figure 5a shows the increase in the divertor $D_\alpha$ signal indicating the start of EDA H-mode phase. The QCM appears with a sharply decreasing frequency starting at ~ 80 kHz in several diagnostics as the $D_\alpha$ signal begins to rise. The figure also shows that the QCM is quasi-coherent initially when its frequency is decreasing (till ~ 2.7 s) and becomes more coherent after that. The QCM frequency keeps decreasing continuously as the discharge progresses, which is likely due to a change in the background plasma parameters as the density and power vary during the discharge. The change in background $E \times B$ velocity can also affect the frequency of the QCM, and it will be shown in later sections.

The QCM is seen clearly in the DBS phase (figure 1h) and amplitude (figure 2b) spectra from DBS60 system. As discussed in section 2, any coherent velocity perturbation of the Doppler shift is detected by DBS phase, while the amplitude spectrum is related to the level of density fluctuation over a weighted wavenumber range. A harmonic of the QCM is seen in DBS phase as well as amplitude spectra. Figure 2c shows the QCM in the spectrum of density fluctuations measured by BES. It is to be noted here that the QCM appears in BES spectrum even before the NBI pulses are introduced at 2.8 s for the BES measurements. The NBI is injected beginning at 2.5 s in pulses until 3.14 s and is held fixed after that. The spectral power of QCM increases with each injection of NBI pulse and remains at a higher level once NBI power is held constant. The observation of QCM without the NBI occurs due to the presence of strong density fluctuations in the background light collected by the BES optics. Similar observations have been reported in Alcator C-Mod tokamak [27]. The localization of the fluctuation in such case is valid only when the NBI beam is turned on [27]. The first harmonic of the QCM is seen in the BES fluctuations spectra. The electromagnetic nature of QCM is confirmed by its presence in the magnetic fluctuation spectrum measured by a magnetic probe (figure 2d). It is interesting to note that the start frequency of the QCM detected with the magnetic probe is similar to the frequency observed in other diagnostics. However, the frequency evolves differently in a way that the ending frequency is twice of that observed in other diagnostics. Although interesting, no consistent explanation is found for this observation. Finally, the detection of QCM in a Langmuir probe current spectrum is shown in figure 2e. The probe is on the divertor shelf (R=1.5 m), at a distance of $\Psi_N$=1.05 ($\Psi_N$ is the square root of the normalized poloidal flux), and ~2 cm away from the outer strike point. The distances are calculated after mapping to the midplane, which means that the plasma is in the far SOL with eelctroj temperature ($T_e$) below 10 eV. Note that the QCM frequencies in each diagnostic are in good agreement with each other, except in the magnetic probes.

*(c) Observation of QCMs at higher-k using DBS*

The Doppler shifted intensity spectrum (complex spectra of I+iQ, where I is the measured in-phase quadrature component and Q is the out of phase quadrature component) for intermediate-k ñ is shown in figure 3 for a single channel from each of the two DBS systems (DBS 60 and DBS 5, which are separated toroidally by 180°) during the EDA H-mode phase. Here: DBS60 system measures the ñ at eight spatial points localized near ρ ~ 0.95 – 0.98 while the measurements of DBS5 system are localized near ρ ~ 0.96. The positive Doppler shifts refer to the propagation of the density fluctuations in the electron diamagnetic drift direction in the lab frame. The spectrum is obtained by windowed short time Fourier transform (STFT) (3.25 ms for each window) method with a 50 % overlap employing Hanning window. A 20 kHz smoothing is performed in the frequency domain. All DBS signals are digitized at 5 MHz.

Figure 3a and figure 3b shows the measurement of intermediate-k ñ at $k_\perp$ ~ 5 – 6 cm$^{-1}$ using the DBS60 system and at slightly lower $k_\perp$ ~ 2 – 3 cm$^{-1}$ using the DBS5 system respectively.

Both the measurements are localized near ρ ~ 0.96 during the EDA H-mode phase. The solid black line corresponds to the Doppler shift of the ñ which is estimated by taking a weighted average of the spectral intensity over the frequency range -200 to 1500 kHz. It can be seen that as the discharge transitions to EDA H-mode regime, the Doppler shift jumps to ~ 600 – 700 kHz for the higher-k (5–6 cm$^{-1}$) measurement and ~ 300 – 400 kHz for the lower-k measurement (2–3 cm$^{-1}$). The different Doppler shifts are consistent with the two channels measuring different wavenumbers (5–6 cm$^{-1}$ and 2–3 cm$^{-1}$) with the same local ExB velocity. It is interesting to note that the mean Doppler shifts for both systems decrease in time in a manner similar to the evolution of the QCM frequency (figure 2b-2e). This can be seen in figure 4 which shows QCM in the phase spectra of the DBS for the same discharge (# 161238) and DBS channel while the solid white line is the mean Doppler shift taken from figure 3a and scaled by a factor of 1/10. The factor 1/10 is chosen such that the mean Doppler shift approximately matches the QCM frequency in the phase spectra. This comparison is helpful to visualize the similar time evolution of the QCM frequency and the mean Doppler shift. It will be shown later that the time evolution of the QCM frequency and the mean Doppler shift is consistent with the evolution of background $E \times B$ velocity.

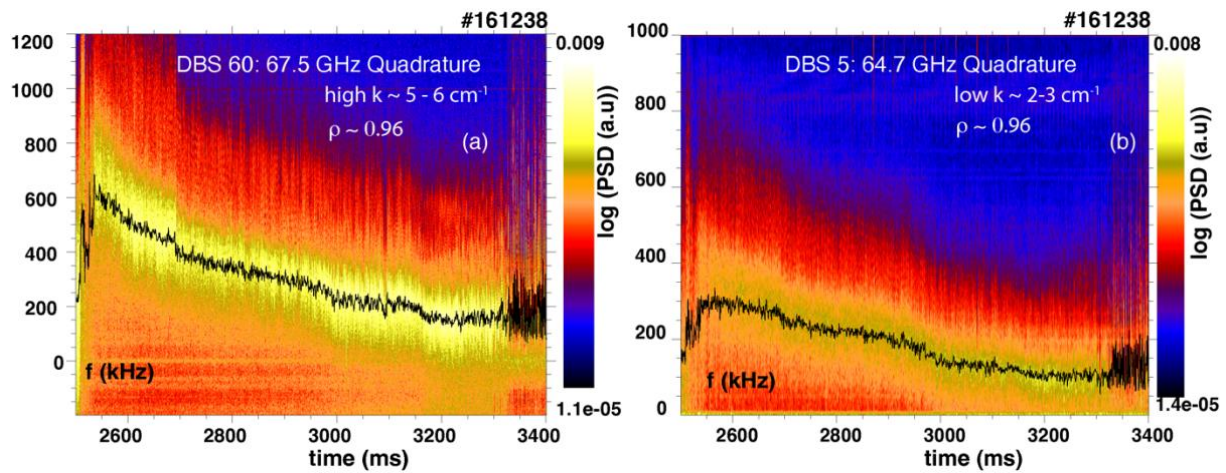

Figure 3: Spectrum of Doppler shifted ñ measured using (a) DBS60 system (X-mode) at higher $k_\perp$ ~ 5 – 6 cm$^{-1}$ and (b) DBS5 system (X-mode) at lower $k_\perp$ ~ 2 – 3 cm$^{-1}$. The solid black line represents the Doppler shift calculated using the weighted average of spectral intensity. The different Doppler shifts are consistent with the two channels measuring different wavenumbers (5–6 cm$^{-1}$ and 2–3 cm$^{-1}$) with the same local E×B velocity.

The intermediate-k ñ ($k_\perp \rho_s$ ~ 1.0) is shown in figure 5b for shot 161235 while the phase spectrum is shown in figure 5a. Here, the shot 161235 is similar to shot 161238 apart from having a lower line-averaged density. Figure 5b is obtained by resolving the Doppler shifted ñ spectrum with proper frequency and time resolution, and reducing the smoothing in the frequency domain. It can be seen that the ñ spectrum consists of multiple, discrete modes with their frequencies decreasing in time similar to that of the QCM frequency observed in figure 5a. The power spectral density of the ñ spectrum at a single time instant ~ 1905 ms is shown in figure 5c. The figure shows that the power spectral density consists of distinct peaks separated by ~ 30 kHz. The frequency of the QCM observed in phase spectrum (figure 5a) at the same time (~ 1905 ms), is also ~ 30 kHz. This observation indicates that the ñ spectrum consists of multiple harmonics of the fundamental QCM, which consequently explains time evolution of the ñ as similar to that of the QCM frequency.

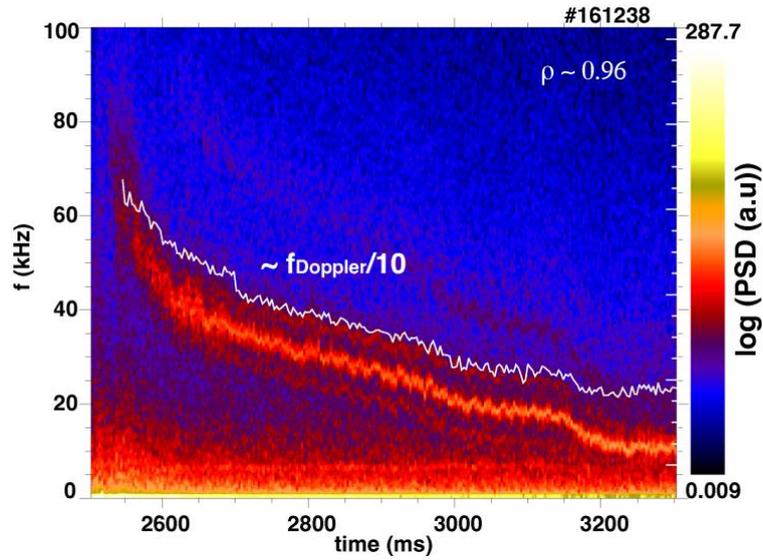

Figure 4: The DBS phase spectra (DBS 60; X-mode) showing the QCM with the Doppler shift (solid white line) obtained from the complex spectra for the same channel and scaled by a factor of 1/10.

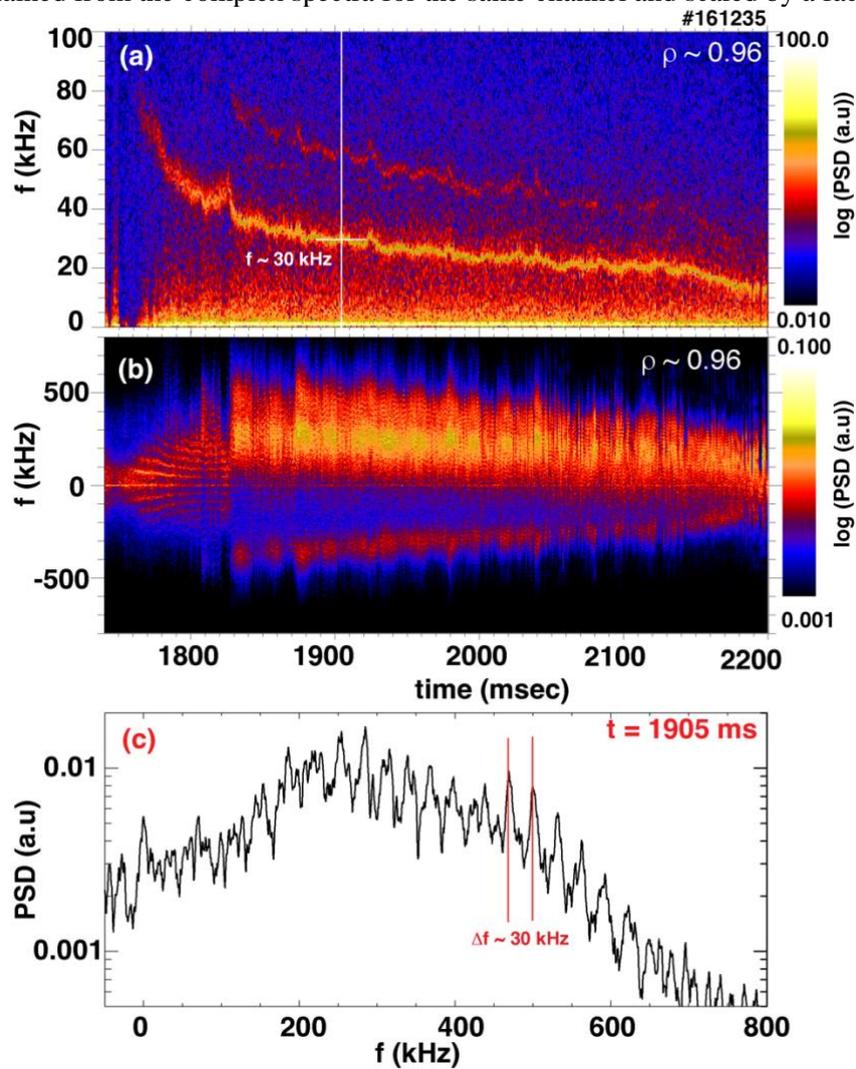

*Figure 5: (a) DBS phase spectra, (b) Doppler shift ñ spectrum and (c) power spectral density for (b) at 1905 ms (marked by vertical, white solid line) using DBS60 in X-mode.*

Figure 5a shows the QCM in the DBS phase spectrum for the same location and channel. As the QCM appears at a frequency of ~ 80 kHz in the phase spectra (figure 5a) at ~ 1760 ms, it is initially seen in the ñ spectrum at a similar frequency. As the frequency of the QCM decreases, higher harmonics of the QCM (up to ~ 10) can be seen in the ñ spectrum up to ~ 1830 ms. At ~ 1830 ms, the ñ spectrum changes abruptly and consists of higher harmonics of the QCM between 5 – 15. The corresponding ñ spectrum for the DBS5 system detects the harmonics between 4 – 7. This could be attributed to the difference in the probed $k_\perp$ values; while the DBS60 probes $k_\perp$ ~ 5 – 6 cm$^{-1}$, the DBS5 probes $k_\perp$ ~ 2 – 3 cm$^{-1}$. Further, as the probed $k_\perp$ decreases towards the LCFS for different channels, the number of harmonics observed in the ñ also decreases. It is important to note that the ñ spectrum across all eight DBS channels consists of both the background turbulence in the range of $k_\perp\rho_s$ ~ 0.1 – 1.2 as well as the harmonics of the QCM co-existing in the same range of $k_\perp\rho_s$. Here, $\rho_s = \sqrt{c_s/\omega_{ci}}$ is the ion sound radius, where $c_s$ is the ion sound speed and $\omega_{ci}$ is the ion cyclotron frequency. However, the interaction of background turbulence and QCM and their relative contributions in the spectrum are not studied in this paper.

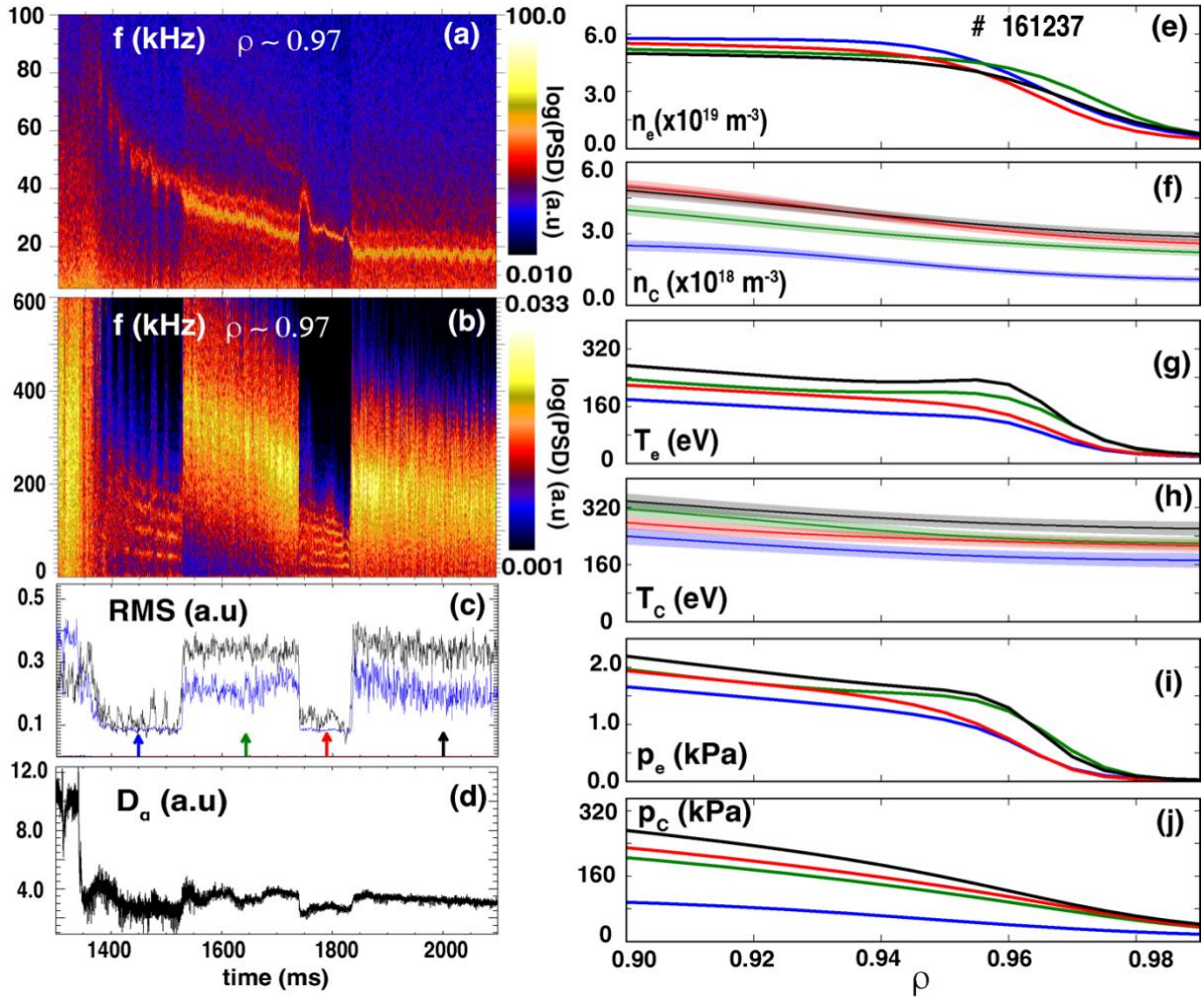

*Figure 6: (a) DBS phase spectrum, (b) Doppler shifted ñ spectrum (DBS60; X-mode), (c) RMS value of the ñ spectrum (blue) evaluated for 10 – 1000 kHz and RMS value of the phase spectrum (black) evaluated for 10 – 100 kHz, and (d) divertor $D_\alpha$ level. The electron density (e), carbon density (f), electron temperature (g), carbon temperature (h), (i) electron pressure profile, and (j) carbon ion pressure profile is calculated at the times 1450 ms, 1640 ms, 1780 ms, and 2000 ms indicated by blue, green, red, and black arrows respectively with data averaged over 50 ms.*

The abrupt shift in the observed harmonics of QCM is analyzed in detail and shown in figure 6 for shot 161237 ($B_T \sim -1.4$ T). Shot 161237 is similar to the shot 161238 shown previously, except that it transitions to EDA H-mode earlier in the discharge. Figure 6a shows the DBS phase spectrum while figure 6b shows the ñ spectrum at the same location. The QCM appears at ~ 1350 ms as the discharge transitions to EDA H-mode. The ñ spectrum (Fig 6b) consists of multiple transitions between lower harmonics to higher harmonics of the QCM. Such transitions are marked by several important changes in the discharge. It is seen that the shift to higher harmonics is accompanied by an increase in the fluctuation amplitude defined as the root mean square value of the ñ in the range of 10 – 1000 kHz (figure 6c) as well as the amplitude of the QCM which is calculated by taking the RMS of the phase spectrum between 10 – 100 kHz (figure 6c). Further, a rise in the divertor $D_\alpha$ levels (figure 6d) is also seen during a shift to higher harmonics. The simultaneous increase in the fluctuation amplitude and the $D_\alpha$ levels is consistent with an increased particle transport across the LCFS during the transition to the higher harmonics. It is interesting to note that the QCM, as seen in the phase spectrum, is more coherent during the low fluctuation amplitude levels and becomes quasi-coherent as the fluctuation amplitude increases. It has been previously reported that the QCM becomes more coherent as the $q_{95}$ decreases towards a value of ~ 3 [27]. However, the $q_{95}$ in this discharge remains near 3.5 during the entire EDA H-mode phase. The profiles of electron density, carbon density, electron temperature, carbon temperature, electron pressure, and carbon ion pressure are plotted in figure 6(e-j) respectively to understand the regimes characterized by low and high fluctuation amplitude levels. The electron profiles are measured using Thomson scattering amd the ion profiles are measured using the CER diagnostics, as discussed in section 2. The profiles are analyzed during the low fluctuation phases (~ 1450 ms depicted by blue line and ~ 1780 ms depicted by red line) and high fluctuation phases (~ 1640 ms depicted by green line and ~ 2000 ms depicted by black line). The profiles are averaged over 50 ms in each phase. It can be seen clearly from figure 6i and 6g that a sharp difference exists in the pedestal electron pressure and electron temperature values between the low and high fluctuation phases. However, the changes in the electron temperature and electron pressure analyzed at pedestal top occur gradually (within ~ few ms) while the transitions in the phase as well as ñ spectra are fast (< ms); the reason for fast transition is not apparent. The low fluctuations, lower harmonics phase (blue, red lines) have lower electron pressure pedestal compared to the high fluctuations, higher harmonics phase. Electron temperature too shows a similar behavior from the separatrix to the outer half of the pedestal. However, the other plasma parameters do not exhibit any clear differences between the two phases. Further, the measurement of distinct harmonics in two regimes is not related to the change that occurs in the radial locations probed by DBS; rather, it represents an actual change in the plasma parameters. Thus, it is worth exploring whether the QCM and its harmonics might be edge localized and electron temperature/pressure gradient driven instabilities. In addition, the abrupt transition between the lower and higher harmonics poses an interesting physics question and will be addressed in future publication.

*(d) Localization of QCM:*

The radial variation of the QCM amplitude during the EDA H-mode phase is measured using the BES diagnostics. Figure 7a shows the normalized relative density fluctuation amplitude (dn/n) of the QCM (depicted with black diamonds) measured by BES. The vertical error-bars in the figure signify the statistical error in the measurement while the horizontal error-bars depict the radial extent of each measurement channel. The fluctuation amplitude is calculated in the range of 5 – 30 kHz for the shot 161238 (figure 1) and averaged within the time range of 3210 – 3310 ms, when then the discharge remains stationary for ~ 100 ms. The figure clearly shows that the QCM is localized in the edge region and peaks just inside the separatrix near $\rho \sim 0.95$. The peak fluctuation level is close to $\sim 40\%$ as seen in the figure. Figure 7b and 7c shows the electron and carbon ion pressure profile respectively along with the individual measurements. The electron profile is measured using Thomson scattering and carbon ion profile is measured using CER. The statistical as well as the fitting error is propagated to calculate the error in the pressure gradients. The peak of QCM amplitude coincides with the maximum of the electron as well as the ion pressure gradients within the measurement uncertainty (gradient are depicted in blue and green respectively with the shaded region indicating the $1\sigma$ error bar), which are obtained at the same time. The radial variation of the QCM amplitude appears to be related to the electron as well carbon ion pressure gradient. This is also consistent with the results presented in the previous section (figure 6i) where the distinct regimes of low and high harmonics were marked by different pedestal electron pressure.

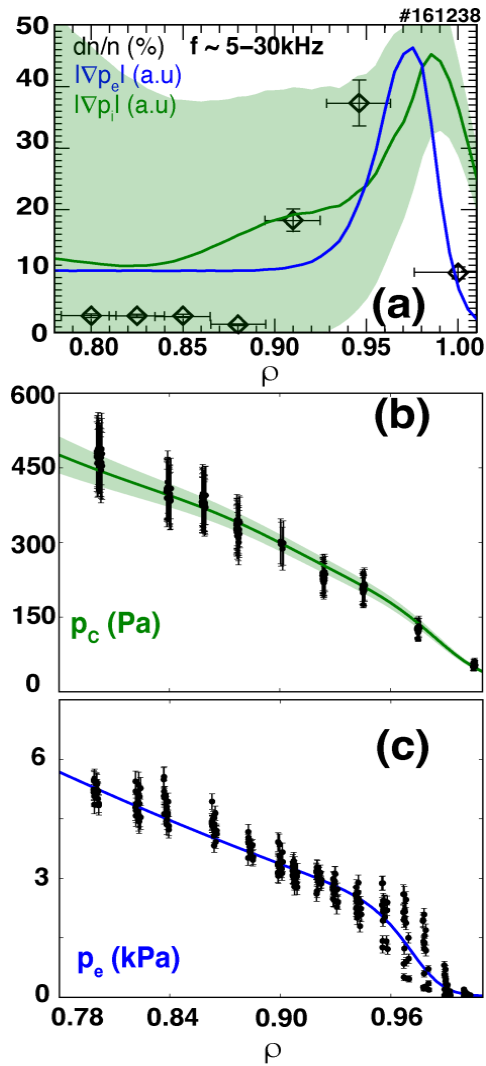

*Figure 7: (a) Radial variation of the normalized density fluctuation amplitude from BES in the range of QCM (5 – 30 kHz) shown with black diamonds. The electron and ion pressure gradients in arbitrary units are shown in blue and green respectively with $1\sigma$ error bar depicted by the shaded areas, (b) electron pressure measured using Thomson scattering, (c) carbon ion pressure measured using CER. The data is shown for shot 161238 for time 3210–3310 ms.*

The measurement using BES show that the peak density fluctuation level is $\sim 40\%$. This raises the question whether the measurement of the harmonics of QCM at higher wavenumber using DBS are real or diagnostic artefacts. DBS measurements are expected to have non-linear response with such high fluctuations due to changes in the path length of the beam and angle of the cutoff layers [46]. To resolve this, we note that BES measures the wavenumber of the fundamental QCM to be ~0.5 cm$^{-1}$, while the DBS measures its higher wavenumber components which fall in the range of 5–6 cm$^{-1}$. According to the wavenumber spectrum for the turbulence, the power falls off as k$^{-3}$ or possibly even more steeply [48–51]. Using a wavenumber power spectral decay of k$^{-3}$, we calculate a power reduction of ~ 0.001. This results in an amplitude decay of 40%*($\sqrt{0.001}$), with the density fluctuation <1.3 % for k~5 cm$^{-1}$ or 0.9% for 6 cm$^{-1}$. This is significantly less than 40%, and pushes the DBS towards the linear response regime. To carry this analysis further, we look at the non-linear effects as reported in Ref. [46]. There, non-linear effects are calculated to be important for DBS when:

$$\gamma \equiv \frac{\omega^2}{c^2} \frac{\tilde{n}^2}{n_c^2} x_c l_{cx} ln \frac{x_c}{l_{cx}} > 1$$

Here, where ñ is the root mean square (rms) of the turbulence amplitude, $n_c$ is the plasma cut-off density, ω is the probing frequency, $x_c$ is the distance from the plasma edge up to the cut-off, $l_{cx}$ is the radial correlation length of the turbulence, and c is the speed of light in vacuum. For an incident microwave beam of 75 GHz, probing wavenumber of 6 cm$^{-1}$, 1% density fluctuation, the distance from plasma edge to cutoff ~ 5 cm, and the radial correlation length ~ 2cm (~10$\rho_i$) which correspond to  γ < 0.1. From this prediction the DBS measurements are well within the linear regime. Thus, this evidence suggests that the harmonics of QCM measured at higher wavenumber are real and not diagnostic artefacts.

The results presented in this section reveals that the QCM consists of a fundamental frequency and its harmonics. These harmonics appear in the intermediate-k ñ spectrum measured using DBS with (k$_\perp \rho_s$ ~ 0.1 – 1.2). Certain discharges consist of distinct EDA H-mode regimes with low and high fluctuation levels which corresponds to lower and higher divertor D$_\alpha$ level. The lower/higher fluctuations phase contain low/high harmonics of QCM (1–5 in lower, 5–15 in higher) and are characterized by distinct pedestal temperature and pressure values. Further, the QCM is shown to exist in the edge (ρ>0.88) and scrape-off layer.

## 4. Stability and transport properties of EDA H-mode discharges

This section focuses on the pedestal stability and transport during the EDA H-mode discharges. A stationary time during the during the discharge is chosen for the study. The plasma parameters change by < 10% during this phase.

Figure 8 shows the EDA H-mode in DIII-D discharges for shot 161231 (B$_T$ = -2.0 T, I$_p$ = 1.0 MA). The EDA H-mode is multiple times in a single discharge for this shot, first after the L – H mode transition (1.9 – 2.8 s) and the other between the ELMy H-mode and L mode transition (3.48 – 3.87 s). Both the phases depicted in the blue shaded boxes are marked by ELM-free period (figure 2b) and appearance of QCM (figure 8d). The second EDA H-mode phase appears when torque is reduced to nearly 0 N-m and the NBI power is decreased to ~ 2 MW. However, the density remains nearly constant (< 4% variation) during the second EDA H-mode phase. Further, the discharge transitions to an ELMy H-mode after the first EDA H-mode phase with an increase in the NBI power and input torque. The QCM disappears as the discharge starts to ELM. After the second EDA H-mode phase, the discharge transitions to the L-mode plasma after a decrease in the NBI power below 1 MW. The observation of EDA H-mode in several discharges as well as multiple transitions of EDA H-mode (from L-mode as well as ELMy H-mode) suggests that it is a reproducible phenomenon.

The ELM-free EDA H-mode exhibits pedestals in both electron density and temperature. Figure 9 shows the kinetic profiles calculated during the stationary phase (2000–2100 ms) of the discharge 161231 (shown in figure 8). Figure 9a (9b) shows the individual data points and the corresponding fits for the electron density (temperature) and the carbon density (temperature). The electron density and temperature are measured using the Thomson scattering diagnostics. These profiles are fitted using the modified tanh fitting [52]. The ion density and temperature profiles are measured using CER diagnostics and fitted with spline fitting. The ion/rotation profiles and electron profiles are radially adjusted according to the criteria discussed in section 2. Figure 9c shows the electron and ion pressure profile, fitted with modified tanh and spline fitting respectively. Since the discharges are NBI-heated and the collisionality is high, the core ion temperature is similar to the electron temperature during the EDA H-mode phase.

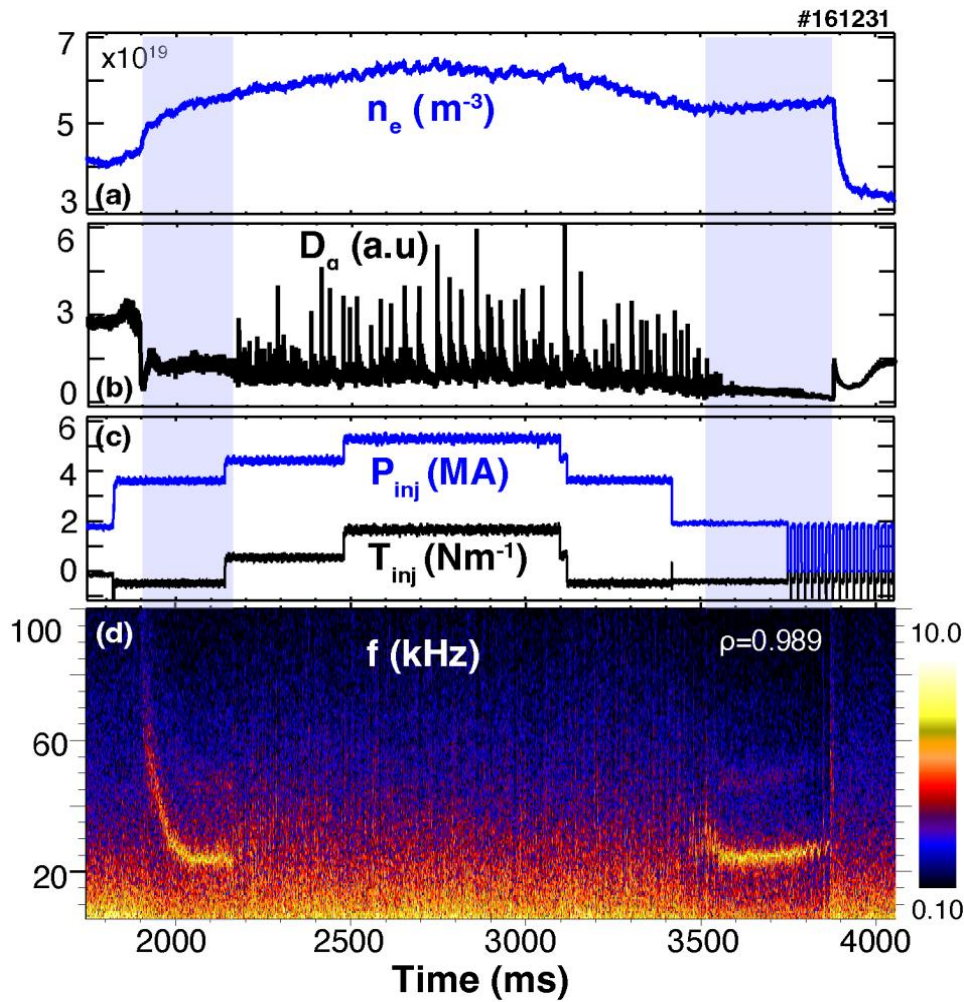

*Figure 8: Multiple EDA H-mode phases during a single discharge. (a) line-averaged density, (b) divertor $D_\alpha$ signal, (c) NBI power (blue) and input torque (black), and (d) DBS phase spectra (DBS8; 62.5 GHz; X-mode). Shaded blue region indicates the EDA H-mode phase.*

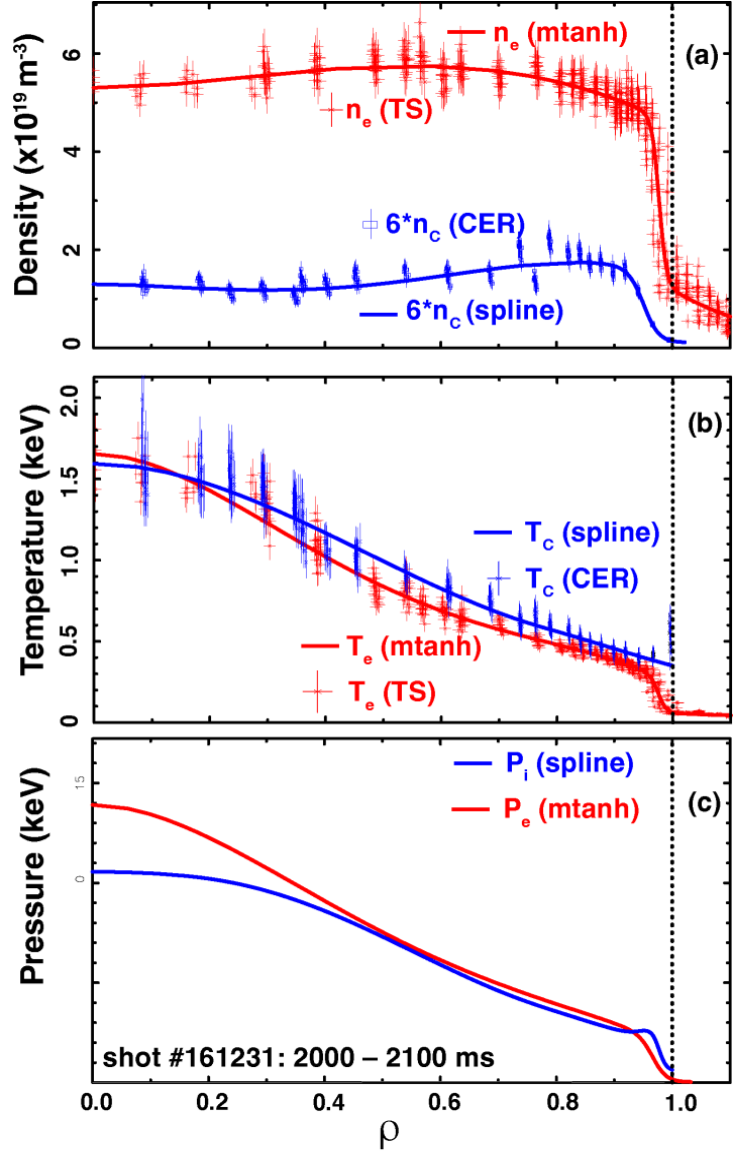

Figure 9: a) Profiles of electron (red) and carbon ion ($C^{6+}$, blue) density with individual data points, (b) profiles of electron (red) and carbon ion ($C^{6+}$, blue) temperature, and (c) profiles of electron (red) and the main ion (blue) pressure for shot 161231.

The pedestal stability to type I ELMs during the EDA H-mode is calculated using the ELITE code [53] with the results shown in figure 10. Figure 10 shows the stability diagram for the pedestal during the stable EDA H-mode phase, calculated for shot 161231 for a time 2000 – 2100 ms (figure 9, input profiles shown in figure 8). To determine the stability boundary shown in the figure 10, a set of equilibria are generated covering a grid in the normalized pressure gradient ($\alpha$) and the normalized pedestal current density ($j_N$) by independently varying the pedestal pressure and current density relative to the experimental kinetic equilibrium. Here, the normalized pedestal current density $j_N$ is $j_{peak}/2\langle j \rangle$. where $j_{peak}$ is the peak current density at the pedestal and the $2\langle j \rangle$ is twice the volume average current density. This variation is performed keeping the total current and stored energy fixed by adjusting the central profiles to compensate for the pedestal changes. The plasma shape is also kept fixed. The pedestal pressure is scaled by a simple multiplier thus keeping the pedestal width fixed. ELITE is then run for a series of toroidal mode numbers ($n$), typically in the range of $n = 5$–$30$, for each equilibrium on the grid. The stability threshold is set to the point where the growth rate exceeds diamagnetic stabilization effects [54], i.e. $\gamma_{max}/(\omega_{eff}/2) < 1$. The experimental point is marked by a crosshair with an open square on the stability diagram, demonstrating that the pedestal lies very close to the ideal ballooning limit within the experimental uncertainties so that a slight increase in the pressure gradient can lead to ELMs. This can be seen from figure 8, where an increase in the NBI power leads to an increased pressure gradient (not shown) and the appearance of ELMs.

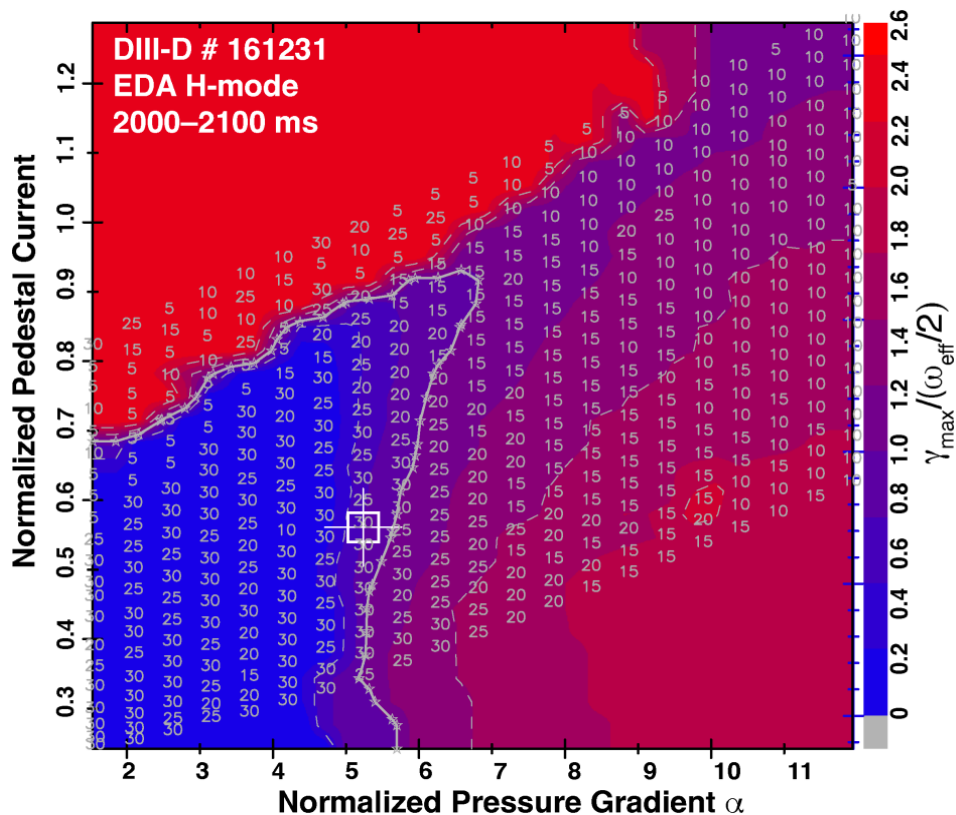

*Figure 10: Pedestal stability in EDA H-mode using ELITE [53]. The experimental point is marked by a cross-hair with an open square on the stability diagram. The solid line represents the peeling–ballooning stability boundary and the integers indicate the most unstable toroidal mode number.*

The measurement presented in previous section using DBS showed that the QCM and its harmonics exist in the intermediate-k ñ spectrum with ($k_\perp \rho_s \sim 0.1 - 1.2$). The Doppler shifted ñ spectrum obtained using DBS provides the propagation velocity ($v_\perp$) of the density turbulence in the lab frame. This $v_\perp$ is the combination of background plasma $E \times B$ velocity ($v_{E \times B}$) and the turbulence phase velocity in the plasma frame ($v_{ph}$) and is given by: $v_\perp = v_{E \times B} + v_{ph}$. One can estimate the turbulence phase velocity in the plasma frame given that $v_{E \times B}$ and $v_\perp$ are known. Here, $v_\perp$ is related to the Doppler shift of the backscattered radiation by density fluctuations (of wavenumber $k_\perp$) and can be expressed as $\omega_{Doppler} = k_\perp v_\perp$. The Doppler shift is obtained by performing fits to the ñ spectrum using a generalized Gaussian function [46]. The background $E \times B$ velocity ($v_{E \times B}$) of the plasma can be determined using the CER diagnostics. Using the measurement of $v_{E\times B}$ from CER and $v_\perp$ from DBS, the QCM phase velocity is seen to be negligible in the plasma frame.

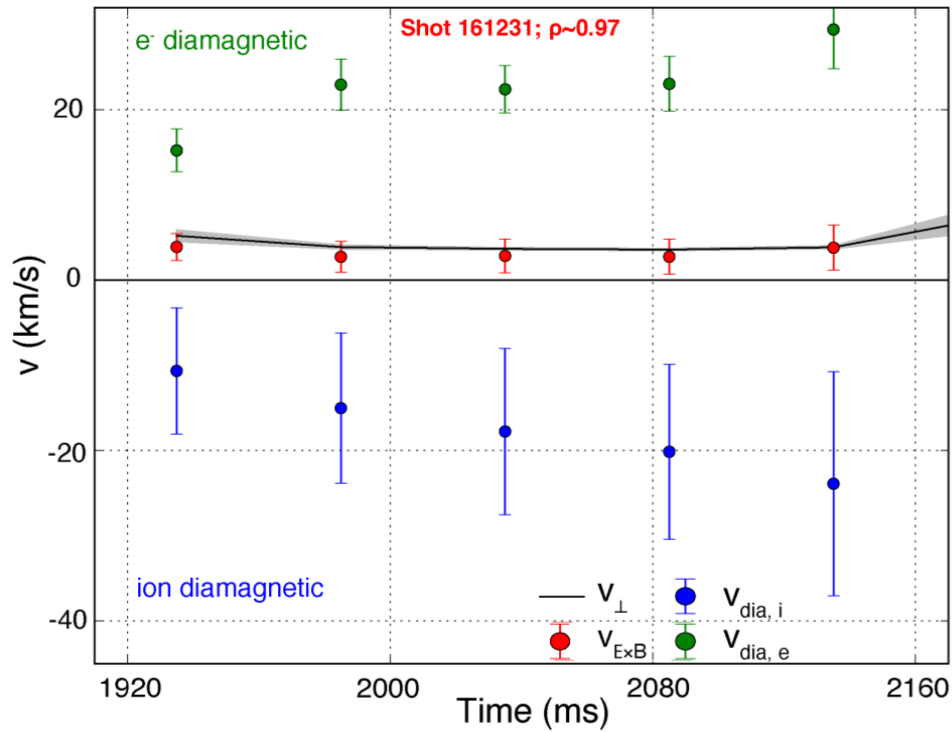

*Figure 11: The velocity (black) of the QCM in lab frame measured with DBS compared with the plasma $E \times B$ velocity (red, from CER), electron diamagnetic velocity (green, from Thomson scattering) and ion diamagnetic velocity (blue, from CER). The positive (negative) velocities represent the electron (ion) diamagnetic direction.*

The above analysis is performed in the case of EDA H-mode to determine the phase velocity of the QCM in plasma frame and the results are shown in figure 11. As shown in previous sections, the Doppler shifted ñ spectrum during EDA H-mode consists of harmonics of the QCM. A generalized Gaussian function fit of the ñ spectrum then provides an estimation of the Doppler shift for a harmonic of the QCM which has a $k_\perp$ most sensitive to the wavenumber of the probed beam. The propagation velocity ($v_\perp$) of the QCM in the lab frame velocity obtained using above analysis is shown with black solid line. Each data point shown in figure 12 is integrated over 50 ms to reduce the uncertainty. For the Doppler shifts, the data is fitted in the frequency space with a Gaussian function every ~ 2 ms in each of the 50 ms interval to obtain the statistical uncertainties. The figure shows that the propagation velocity of the QCM in lab frame agrees well within the statistical uncertainties with the plasma $E \times B$ velocity ($v_{E\times B}$, shown in red) obtained from CER. Further, the electron ($v_{dia,e}$) and ion ($v_{dia,i}$) diamagnetic drift velocities are shown in green and blue respectively. It can be seen that the magnitudes of $v_{dia,e}$ and $v_{dia,i}$ are much higher than the plasma $E \times B$ velocity. Hence, the phase velocity of the QCM in plasma frame is inferred to be very small and consequently it predominantly rotates in the electron diamagnetic direction in the lab frame. The initial decrease in the QCM frequency (from ~ 80 kHz to 20–30 kHz as shown in figure 8d) can also be explained as a consequence of the initial decrease in the magnitude of the radial electric field ($E_r$). The increase in the ion density and toroidal rotation in the co-$I_p$ direction offsets any increase in the ion pressure gradient, leading to a decrease in $E_r$ and consequently in the plasma $E \times B$ velocity.

These results thus present the time–resolved measurements of electron and ion density, temperature, plasma rotation and radial electric field during the EDA H-mode phase and explains the full dynamics of the QCM. We infer the QCM to be a separatrix spanning mode, peaking just inside the separatrix and closely following the electron pressure gradient, existing in a wide range of k$_\perp$ρ$_s$ ~ 0.1 – 1.2 with multiple harmonics, and with a negligible phase velocity in the plasma frame.

## 5. Linear CGYRO simulation of EDA H-mode discharges

In this section, we present results from a linear gyrokinetic simulation of EDA H-mode discharge (#161231; figure 8). Simulations are performed in the pedestal region using the CGYRO code [55], using an equilibrium (EFIT) corresponding to 2000 – 2100 ms. The profiles are averaged over this time window (Figure 9) and a corresponding kinetic equilibrium (KEFIT) is subsequently constructed [56]. Simulations are performed at radial locations between $0.945 \leq \rho \leq 0.990$, where the QCM is observed in the experiments. The binormal wavenumber range k$_\perp$ρ$_i$ ~ 0.05–1.0 is chosen to match the experimentally measured range of QCM. Here, ρ$_i$ = c$_s$/Ω$_d$, where $c_s = \sqrt{T_e/m_d}$, Ω$_d$ is the deuterium cyclotron frequency, and $m_d$ is the deuterium mass. We choose to present results from a single ballooning wavenumber $\theta_0 = k_x/k_\perp s = 0$, where k$_x$ is defined as the radial wavenumber evaluated at the outboard midplane and s is the global magnetic shear. Near the end of this section, we discuss modes with ballooning wavenumber θ$_0$ ≠ 0 and their resilience to ExB flow shear.

CGYRO is a local delta-f flux-tube code [57], which solves the ρ$_i$/a = 0 limit of the gyrokinetic equation assuming that local field perturbations are much smaller than the equilibrium fields. While CGYRO has an option to include shear in the profile gradients [58], we do not use them here. One necessary condition for the local limit to be valid is that relevant properties of the equilibrium do not change substantially over the mode's radial extent: k$_r$ L$_{Ts}$ $\gg 2\pi$, where k$_r$ is a mode's radial wavenumber defined as $k_r = k_{perp} \cdot \hat{r}$, and $L_{Ts}^{-1} = -dln(T_s)/dr$, where s is the species label. We show that the dominant modes we find easily satisfy k$_r$ L$_{Ts}$ $\gg 2\pi$. Here, k$_{perp}$ is the perpendicular wavenumber and is defined as $\boldsymbol{k}_{perp} = k_r \cdot \hat{r} + k_\perp \cdot \widehat{k_\perp}$.

Linear gyrokinetic simulations are carried out at multiple radial locations within the pedestal to determine the dominant modes, their growth rates, real frequencies, quasilinear particle diffusivities, quasilinear thermal conductivities and quasilinear fluxes. Simulations are fully electromagnetic, collisional, and three species (electron, ion and carbon impurity) are treated kinetically. While we have used a high-fidelity Sugama collision operator [59] we found that our simulation results were very similar with collisionality reduced by a factor of 100. The mode type is deduced using a fingerprints-like analysis [60], using the ratio of electron particle diffusivity to thermal conductivity, the ratio of electron to ion heat fluxes, the mode parity, and the sensitivity to equilibrium gradients. Figures 12 – 16 show that two dominant modes exist at different radial locations. From $0.945 \leq \rho \leq 0.958$, the dominant mode shares characteristics most closely associated with trapped electron modes (TEM) [61, 62]. TEM pedestal turbulence has been observed experimentally in NSTX [63] and in linear and nonlinear gyrokinetics simulations [62, 64, 65, 66]. For $\rho > 0.962$, the toroidal branch of electron temperature gradient (ETG) modes dominates [67, 68, 69]. ETG pedestal turbulence has also been observed in numerical work across multiple devices [70, 71, 72, 73], and reduced models have been proposed for ETG pedestal transport [65, 74, 75]. A table summarizing the fingerprints-like analysis [60] to determine the dominant mode type in the simulations (figure 12–16) is shown in table 1. While nonlinear simulations are outside the scope of this study, they would be useful to give more confidence in how and whether TEM and ETG modes give rise to saturated turbulence and its associated transport fluxes.

Table 1: Fingerprints-like analysis to obtain dominant mode type

|  | TEM Theoretical | ETG Theoretical | TEM Simulation ($0.945 \leq \rho \leq 0.958$) | ETG Simulation $\rho > 0.962$ |
|---|---|---|---|---|
| $D_e/\chi_e$ | ~1 | << 1 | ~ 0.1–1 | ~ $10^{-1}$–$10^{-3}$ |
| $D_i/\chi_i$ | ~ 1 | Not used | ~ 1 (not shown) | Not used |
| $Q_i/Q_e$ | ~ 1 | << 1 | ~ 0.1–1 | ~ $10^{-2}$–$10^{-3}$ |
| $\chi_i/\chi_e$ | ~1 | << 1 | ~ 1–10 | ~ $10^{-1}$–$10^{-3}$ |
| Change in γ with increasing β | Decrease | Not Used | Decrease | Not used |

We now discuss the TEM and ETG mode properties in some detail. Figure 12a shows the normalized growth rate of the most unstable mode and figure 12b shows its normalized real frequency as a function of $k_\perp \rho_i$, calculated at various ρ values. Here, the growth rates γ and real frequencies $\omega_R$ are normalized to $c_s/a$, where a is the minor radius. Figure 12 shows two distinct modes, with a transition at $\rho = 0.962$. The TEM has fairly similar growth rates across radii where it dominates, while the ETG exhibits a strong dependence of γ with the radial location. The increase in the growth rate of ETG is consistent with the increase in $a/L_{Te}$ as shown in figure 17. Further, Figure 17 demonstrates that $\eta_e > 1$ (where $\eta_e \equiv L_{ne}/L_{Te}$) across the ρ values we examine, as well as a sharp increase in $L_{Te}$ beyond $\rho \sim 0.962$, both of which are consistent with ETG instabilities [69]. Since the ETG growth rate is much larger than the TEM, it is likely that the TEM is a subdominant mode beyond radial locations $\rho > 0.962$, and hence might play an important role in particle and heat transport at these radii.

While both dominant modes propagate in the electron diamagnetic direction, the TEM's $\omega_R$ value is close to zero, shown in Figure 12b. Here, negative $\omega_R$ values indicate the mode propagation in electron diamagnetic direction. Figure 13b shows that both the modes exist over a wide range of toroidal mode numbers (n). The parity of the mode can be defined as [76] $1 - \frac{|\int A_\parallel d\theta_b|}{\int |A_\parallel| d\theta_b}$ where $A_\parallel$ is the parallel magnetic potential and $\theta_b$ is the extended ballooning angle [77], which is similar to the poloidal angle $\theta$, but satisfies $-\infty < \theta_b < \infty$, and includes information about the change in the radial wavenumber due to the effects of magnetic shear. We plot the mode parity in figure 13a. The TEM modes have a twisting parity while ETG modes exhibit a parity varying from 0.4–0.8, suggesting that magnetic reconnection is present. Tearing ETG has been reported in other simulations [78]. In Figure 14a and 14b, we plot eigenmodes for the TEM and ETG modes across radii for $k_\perp \rho_i$ =0.26. The trapped electron mode amplitudes are centered around $\theta_b = 0$, and can be extended in $\theta_b$, with significant amplitudes for $|\theta_b| \leq 2\pi$. We can estimate the goodness of the local radial approximation ($k_r L_{Ts}$)$\gg 2\pi$) using the radial wavenumber $k_r \rho_i \sim k_\perp \rho_i s \theta_b$, giving $k_r L_{Te} \simeq k_\perp \rho_i s \theta_b (L_{Te}/a)(a/\rho_i)$. For $\rho = 0.958$, $s \simeq 6$, $(L_{Te}/a) \simeq 1/27$, and $\left(\frac{a}{\rho_i}\right) \simeq 1.5 \times 10^3$. For the TEM in Figure 14a, $k_\perp \rho_i$ =0.26 and $\theta_b \simeq \pi$, giving $k_r L_{Te} \simeq 260$, and hence the local approximation is well-posed for these TEM instabilities.

The toroidal ETG modes are localized far along the $\theta_b$ coordinate, showing they have a toroidal nature. As found in [70, 72, 79], toroidal ETG modes in steep temperature gradients are typically highly unstable at large $|\theta_b|$ values because the magnetic drift frequency, which increases with $|\theta_b|$, needs to be sufficiently large to be comparable in size to the fast ExB drifts in steep gradient regions. This results in $k_{perp} \gg k_\perp$, such that even ETG modes at $k_\perp \rho_i = 0.1$ satisfy $k_{perp}\rho_e \sim 1$. Such a large perpendicular wavenumber, $k_{perp}\rho_e \sim 1$, means that the perturbed ion kinetic distribution function is very small, and thus we expect negligible ion heat flux from such modes. Using a similar estimate for $k_r L_{Te}$ as we did above for the TEM, for the toroidal ETG modes with the maximum growth rate at ρ= 0.975, we find $k_r L_{Te} \simeq 970$ ($s \simeq 15$, $(L_{Te}/a) \simeq 1/102$, and $\left(\frac{a}{\rho_i}\right) \simeq 2.5 \times 10^3$). While the toroidal ETG branch has the highest linear growth rate, it is also likely that there is a subdominant slab ETG branch with different transport characteristics [73].

The TEM and ETG modes exhibit very different transport characteristics. The transition between the two modes can be seen in figures 15 and 16. The modes for ρ<0.962 exhibit high particle transport, as evident from $D_e/\chi_e \sim 1$ (figure 15a), comparable ion and electron heat conductivity $\chi_i/\chi_e \sim 1$ (figure 15b) and heat flux $Q_i/Q_e \sim 1$ (figure 16a), and mostly electrostatic heat flux $Q_{ES,e}/Q_{EM,e} \sim 10 - 10^3$ (figure 16b). These characteristics conform closely with those of the trapped electron mode (TEM) turbulence [60]. On the other hand, the ETG modes at ρ>0.962 exhibit weak particle transport as $D_e/\chi_e \sim 10^{-1} - 10^{-3}$ (figure 15a), and mostly electron heat transport with $\chi_i/\chi_e \sim 10^{-1} - 10^{-3}$ and $Q_i/Q_e \sim 10^{-1} - 10^{-3}$ (figure 15b and figure 16a), and are highly electrostatic as $Q_{ES,e}/Q_{EM,e} \sim 10^2 - 10^5$ (figure 16b), despite them having electromagnetic character [80] due to tearing.

We now show that both TEM and ETG modes are resilient to ExB flow shear by performing a scan in the ballooning wavenumber $\theta_0 = k_x/k_y s$. As an approximation, the flow shear can only suppress a mode if $\gamma_E > \gamma s \Delta\theta_0$ [79], where $\gamma_E = \left(\frac{\rho}{q}\right)\left(\frac{d\Omega}{d\rho}\right)$ is the flow shearing rate, $q$ is the safety factory, $\gamma$ is the typical linear growth rate, and $\Delta\theta_0$ is the range of $\theta_0$ values for which the mode is unstable. Considering the flux surface $\rho = 0.958$ with global magnetic shear $s \approx 4$, the TEM has a typical growth rate of $\gamma \approx 0.15\ c_s/a$ and a range of unstable ballooning wavenumbers $\Delta\theta_0 \approx 1$, giving $\gamma s \Delta\theta_0 \approx 0.6$. Hence, we would require $\gamma_E > 0.6\ c_s/a$ to suppress the TEM. However, this flux surface only has a flow shearing rate $\gamma_E \approx 0.1\ c_s/a$, and thus the flow shear rate is too low by roughly an order of magnitude to suppress the TEM linearly. Considering a flux surface where ETG dominates $\rho = 0.975$ with global magnetic shear $s \approx 10$, the typical growth rate is $\gamma \approx 5 - 10\ c_s/a$ and $\Delta\theta_0 \gg 1$. Thus to suppress the ETG mode, we would require a flow shearing rate of $\gamma_E \gg 50 - 100\ c_s/a$, which is orders of magnitude higher than obtainable in tokamak pedestals. Hence, we expect that the flow shear is unable to suppress the TEM and ETG modes linearly.

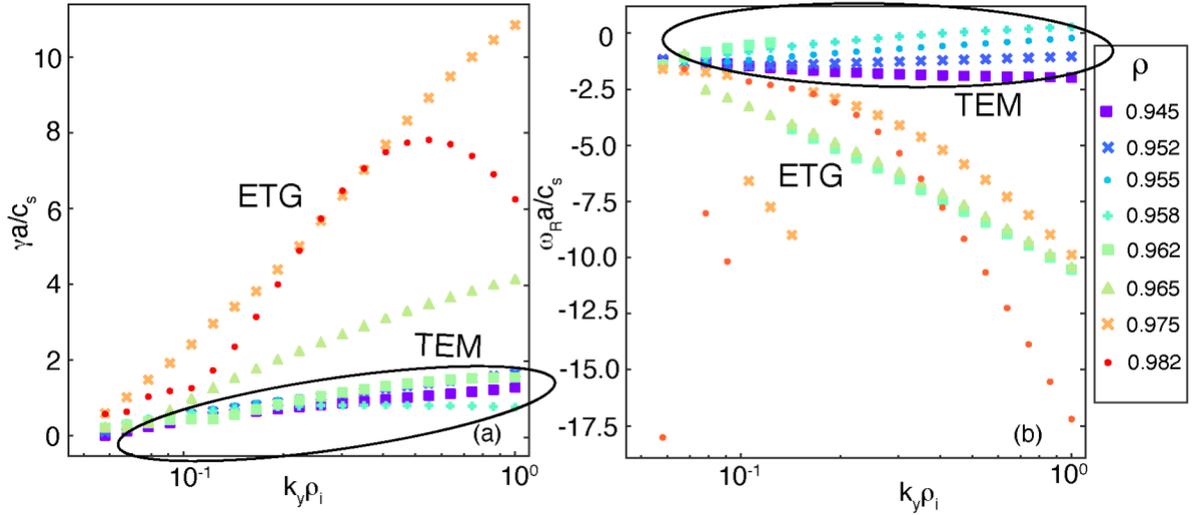

*Figure 11: From CGYRO: (a) normalized growth rate and (b) normalized real frequency with $k_\perp \rho_s$ at different radial locations.*

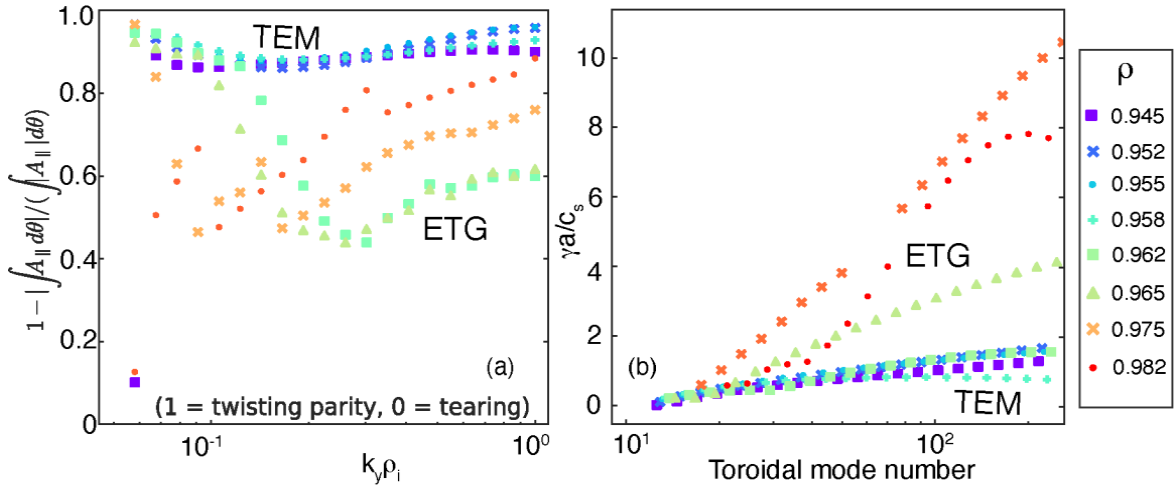

*Figure 13: From CGYRO: (a) Parity of the modes (b) Variation of growth rate with toroidal mode number at different radial locations.*

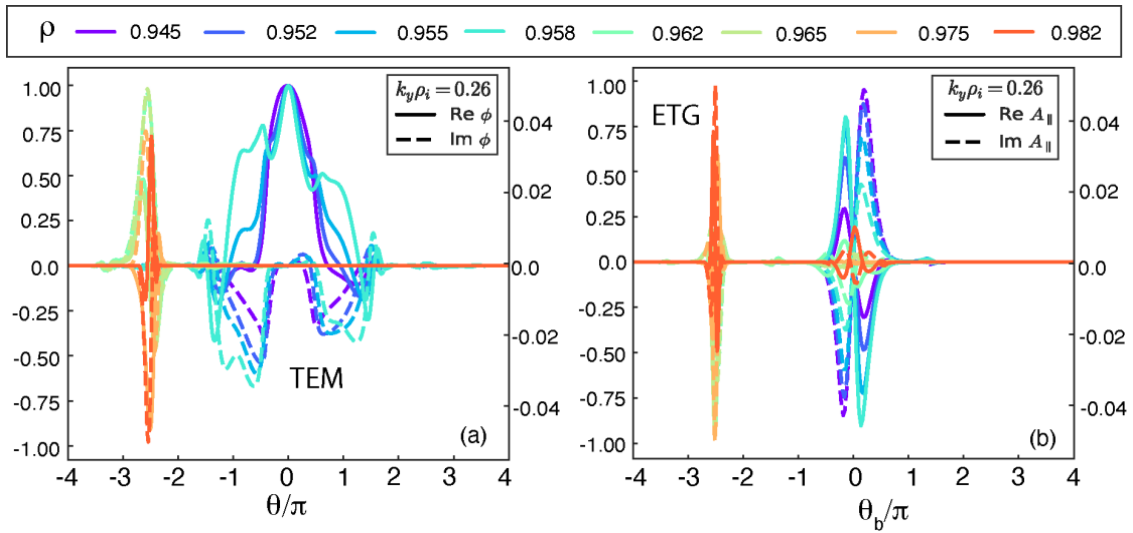

*Figure 12: From CGYRO: Eigenmodes of perturbed electrostatic potential (a) and parallel vector potential (b) at differential radial locations for $k_\perp \rho_i = 0.26$ for different $\theta/\pi$.*

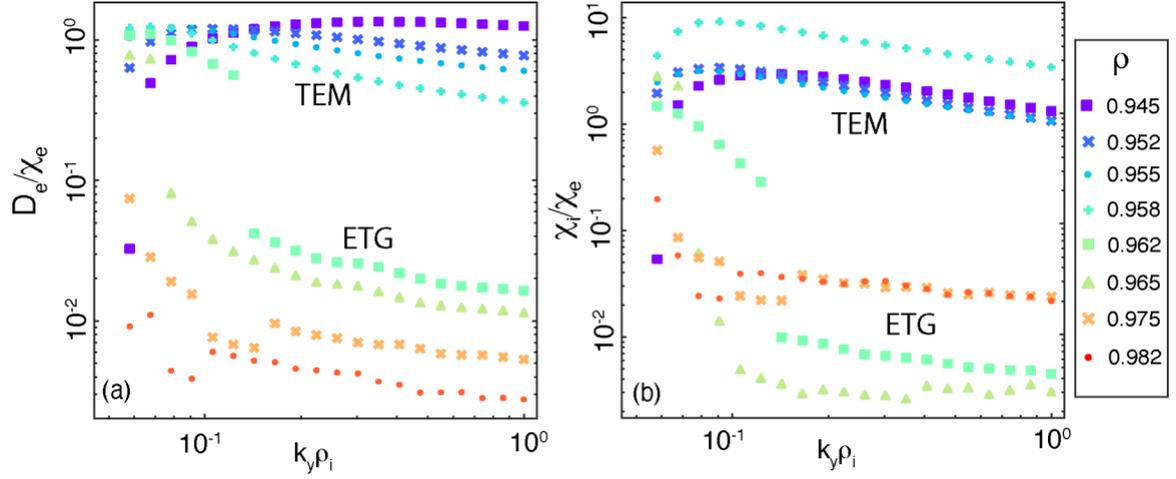

*Figure 13: From CGYRO: (a) ratio of electron particle diffusivity to electron thermal conductivity (b) ratio of electron to ion thermal conductivity with $k_\perp \rho_s$ at different radial locations.*

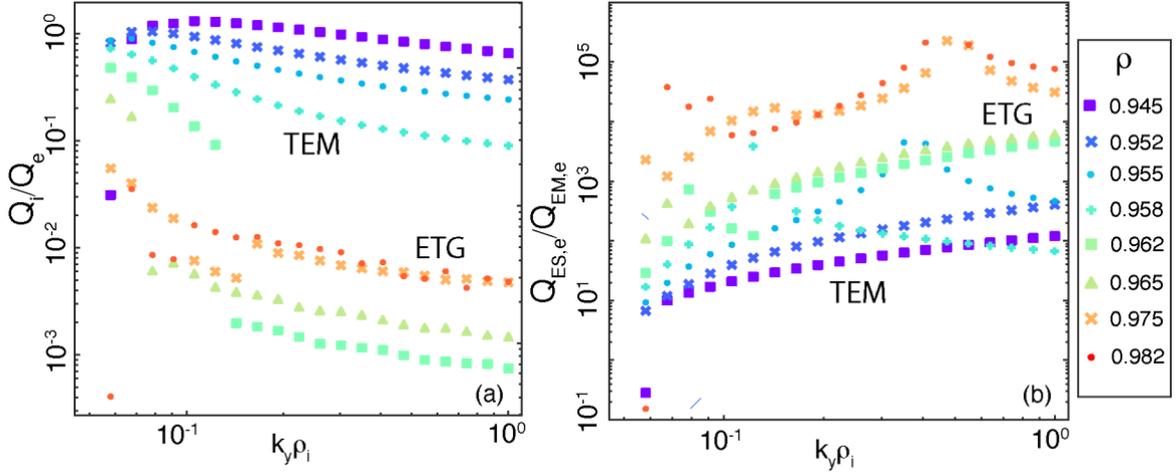

*Figure 14: From CGYRO: (a) ratio of electron to ion heat flux (b) ratio of electrostatic to electromagnetic electron heat flux with $k_\perp \rho_s$ at different radial locations.*

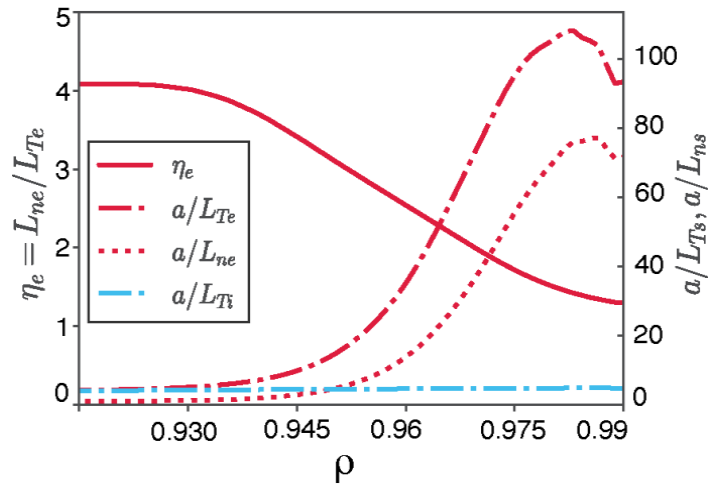

*Figure 15: $\rho$ vs $\eta_e$ on the left-hand side and $a/L_{Ts}$, $a/L_{ns}$ on the right-hand side, where the parameter $\eta_e$ is defined as $\eta_e \equiv L_n/L_{Te}$*

In summary, linear local gyrokinetics simulations of EDA H-mode discharge predict that two dominant modes exist in the same radial and $k_\perp \rho_s$ range as the experimentally observed QCM.

The characteristics of the TEM match quite well with the following experimental observations of QCM. The TEM exists in simulations over a range of $k_\perp\rho_s$ for which the QCM is observed in the DBS. The real frequencies of the TEM are mostly in the electron diamagnetic direction but close to the zero, while in experiments, the rotation of QCM in lab frame is seen to be close to the plasma $E \times B$ velocity indicating that the real rotation is close to zero. In the experiments, the discharge is stable to the coupled peeling-ballooning mode, which suggests that an alternate mechanism exists for the heat and particle transport which maintains an ELM-free pedestal. Several experiments have shown that the appearance of QCM is consistent with an increase in the particle transport. Notably, a decrease in the particle confinement time has been shown with the occurrence of EDA H-mode in Alcator C-Mod [11]. Recent results from ASDEX-U indicate that there is a decrease in the core tungsten impurity during the EDA H-mode [22]. In this experiment, the detection of QCM in the Langmuir probes installed on the lower divertor shelf also indicates the role of QCM in particle transport. The TEMs observed in the simulation of the EDA H-mode in this experiment also cause substantial particle transport ($D_e/\chi_e \sim 1$) and a comparable heat transport in both the electron and ion channels ($\chi_i/\chi_e \sim 1$, $Q_i/Q_e \sim 1$). The experiments show an electromagnetic nature of the QCM while the TEMs are seen to cause non-negligible heat transport in the electromagnetic channel ($Q_{ES}/Q_{EM}\sim 10$) at low value of $k_\perp\rho_s$. Finally, the QCM are seen to be separatrix spanning modes in the experiments. The TEMs are dominant only until $\rho = 0.962$ after which the ETG becomes the dominant instability. However, it is likely that TEMs remain a subdominant instability beyond $\rho = 0.962$. Thus, our linear gyrokinetic simulations reproduce several experimental trends and suggest that the QCM existing during EDA H-mode discharge might be a trapped electron mode instability.

## 6. Summary

EDA H-mode is observed in plasmas in lower single null plasmas with near zero net torque using NBI heating in DIII-D tokamak. EDA H-mode discharges in DIII-D possess several characteristics which are of interest for the future devices such as: good plasma beta and energy confinement (normalized beta $\beta_N \sim 2$, $H_{98y2} \sim 1$), Greenwald fraction ~ 0.75, access conditions very close to the scaled L-H threshold power required for ITER, low impurity content, and no ELMs. Multiple EDA H-mode transitions are observed in a single discharge which suggests that it is a reproducible phenomenon. The EDA H-mode possesses pedestals in both electron density and temperature. The MHD stability calculations for the EDA H-mode show that the operating point is very close to the threshold for the peeling-ballooning instability, but the plasma remains ELM-free. QCM is observed in the density fluctuations measured with BES and DBS, probe current fluctuations of the Langmuir probes as well as in the magnetic fluctuations measured with magnetic probes. DBS, employed to measure QCM, reveals that the intermediate-k ñ spectrum consists of multiple, discrete modes with their frequencies decreasing in time similar to the QCM frequency as well as the inter-mode separation equal to the QCM frequency. Detailed analysis reveals that these modes are multiples of the QCM existing at intermediate wavenumbers ($k_\perp\rho_s \sim 0.1 - 1.2$). Certain discharges consist of distinct EDA H-mode regimes with low and high fluctuation levels which corresponds to lower and higher divertor $D_\alpha$ level. The lower/higher fluctuations phase contain low/high harmonics of QCM (1–5 in lower, 5–15 in higher) and are characterized by distinct pedestal temperature and pressure values.

The QCM amplitude is found to follow the variation in electron as well as carbon ion pressure gradient within the experimental uncertainties and the maximum in QCM amplitude is localized near maximum of the pressure gradient. The propagation velocity of QCM in plasma frame is determined using the Doppler shifted ñ spectrum of DBS which provides the turbulence propagation velocity ($v_\perp$) in the lab frame and the $v_{E\times B}$ velocity measured with the CER diagnostics. This analysis shows that the QCM propagates with the a very small velocity in plasma frame.

It is important to note that the nature of QCM is widely debated in the literature. Several experiments as well as modeling have shown that QCM might be a x-point resistive ballooning mode [25], electron drift wave with interchange and electromagnetic contributions [17], pressure driven surface waves [29] and drift-Alfven wave [28]. However, no single study is able to explain the full physics of the QCM. In this study, the experimentally measured information including the localization of the QCM as well as its intermediate-k structure is employed to perform linear gyrokinetic simulations using CGYRO at several locations within the pedestal. Linear gyrokinetic analysis reveals that from $0.945 \leq \rho \leq 0.958$, the dominant mode shares characteristics most closely associated with trapped electron modes while for $\rho > 0.962$, the toroidal branch of electron temperature gradient (ETG) modes dominate. The characteristics of the TEM match reasonably well with the experimental observation of QCM such as: substantial particle transport ($D_e/\chi_e \sim 1$) and an electromagnetic nature of the QCM at low wavenumber. Further, the TEM is seen to cause comparable heat transport in both the electron and ion channels ($\chi_i/\chi_e \sim 1$, $Q_i/Q_e \sim 1$). It is likely that TEMs remain a subdominant instability beyond $\rho = 0.962$ where ETG dominates. Thus, the linear gyrokinetic simulations reproduce several experimental trends and suggest that the QCM existing during EDA H-mode discharge might be a trapped electron mode instability.

## Acknowledgements

This material is based upon work supported by the U.S. Department of Energy, Office of Science, Office of Fusion Energy Sciences, using the DIII-D National Fusion Facility, a DOE Office of Science user facility, under Award(s) DE-SC0019352, DE-SC0022563, DE-FC02-04ER54698, DE-AC02-09CH11466, and DE-FG02-08ER54999. Part of the data analysis was performed using the OMFIT integrated modeling framework [81].